\documentstyle[prl,aps]{revtex}

\begin{document}
\def\aa{\mbox{\boldmath $a$}}
\def\th{\mbox{th}}
\def\lll{\mbox{\boldmath $\ell$}}
\def\lla{\mbox{\boldmath $\ell_1$}}
\def\llb{\mbox{\boldmath $\ell_2$}}
\def\qq{\mbox{\boldmath $q$}}
\newcommand{\bml}[1]{\mbox{{\boldmath$#1$}}}

\draft
\input{psfig.sty}
\twocolumn[\hsize\textwidth\columnwidth\hsize\csname@twocolumnfalse\endcsname

\title{Deterministic diffusion in flower shape billiards}
\author{Takahisa Harayama}
\address{ATR Adaptive Communications Research Laboratories,\\
2-2 Hikaridai, Seika-cho, Soraku-gun, Kyoto 619-02, Japan}
\author{Rainer Klages}
\address{Max Planck Institute for Physics of Complex Systems, \\  
N\"{o}thnitzer Str.\ 38, D-01187 Dresden, Germany}
\author{Pierre Gaspard}
\address{Center for Nonlinear Phenomena and Complex Systems,
Universit\'e Libre de Bruxelles,\\ 
Campus Plaine, Code Postal 231, Boulevard du Triomphe, B-1050 Brussels, Belgium}
\date{\today}

\maketitle

\begin{abstract}
We propose a flower shape billiard in order to study the irregular parameter
dependence of chaotic normal diffusion. Our model is an open system consisting
of periodically distributed obstacles of flower shape, and it is strongly
chaotic for almost all parameter values. 
We compute the parameter dependent diffusion coefficient of this
model from computer simulations and analyze its functional form by different
schemes all generalizing the simple random walk approximation of Machta and
Zwanzig. The improved methods we use are based either on heuristic
higher-order corrections to the simple random walk model, on lattice gas
simulation methods, or they start from a suitable Green-Kubo formula for
diffusion. We show that dynamical correlations, or memory effects, are of
crucial importance to reproduce the precise parameter dependence of the
diffusion coefficent.
\end{abstract}

\narrowtext
\vskip1pc]

\draft
%
%
%
%
%
%
%
%

\section{Introduction}
\label{intro}

One of the central themes in the theory of nonequilibrium statistical
mechanics is to assess the importance of deterministic chaos for understanding
transport processes such as diffusion \cite{GaspardCUP,Do99}. Simple model
systems appear to be most suited for studying the detailed relation between
microscopic chaos and macroscopic transport. Along this line of research, the
parameter dependent diffusion coefficients of strongly chaotic dynamical
systems have been investigated for one- and two-dimensional mappings
\cite{RechWhite,DMP89,KlagesDorfman,Klages,GaKl}, periodic
Lorentz gases \cite{KlagesDellago}, and billiards in an external field
\cite{HarayamaGaspard}. That diffusion coefficients can be fractal
functions of control parameters was first observed in a simple one-dimensional
mapping generalizing a random walk on the line \cite{KlagesDorfman}. The
origin of this fractality may be attributed to the topological instability of
orbits under parameter variation that affects the parameter dependence of the
diffusion coefficient in a nontrivial way. Based on the analysis of such
simple systems, it was conjectured that fractal diffusion coefficients are
rather generic for low-dimensional fully chaotic dynamical systems exhibiting
some spatial periodicity
\cite{KlagesDorfman,Klages}.  Indeed, recently it was found that in case of
billiards in an external field the diffusion coefficient again exhibits a
highly irregular structure \cite{HarayamaGaspard}. 

The standard periodic Lorentz gas is one of the typical models 
for studying deterministic normal diffusion 
(see, e.g., Refs.\ \cite{GaspardCUP,Do99} and further references therein). 
That it is strongly chaotic and exhibits normal diffusion
was proven by Bunimovich and Sinai \cite{SinaiBunim}. Machta and Zwanzig have
calculated the diffusion coefficient of this model from computer simulations
at some parameter values, and they have matched their results to a simple
analytical random walk approximation \cite{MZ}. That the diffusion coefficient
in the standard periodic Lorentz gas is indeed a non-trivial function of 
the parameter was first reported in Ref.\ \cite{KlagesDellago}. 
Here the analysis by Machta
and Zwanzig was refined by suggesting two methods for systematically
correcting their random walk approximation. However, whether the numerically
detected irregularities in the diffusion coefficient were of a fractal nature
remains an open question. More recently, a third approximation scheme was
proposed by deriving a Green-Kubo formula that exactly generalizes the
Machta-Zwanzig approximation \cite{KlaKor}. Applying all these methods led to
the conclusion that including long-term correlations, or memory effects, was
indispensable to reproduce the precise functional form of the parameter
dependent diffusion coefficient for the standard periodic Lorentz gas.

One of the essential problems in the analysis of diffusion 
in the standard periodic Lorentz gas is
that the parameter range of normal diffusion is very limited.  In this small
region, the irregular behavior of the parameter dependent diffusion
coefficient shows up on very fine scales and appears to be rather smooth
within the range of precision available from computer simulations
\cite{KlagesDellago,TH01}. Consequently, the question about the existence of a
fractal diffusion coefficient is very difficult to answer for this model. As
the main reason for this behavior it might be suspected that the topological
instability of the standard periodic Lorentz gas 
under parameter variation is not strong enough to generate more pronounced 
irregularities in this region. The main purpose of this paper is 
therefore to propose a billiard without external field 
which is very similar to the standard periodic Lorentz gas, but which has a
geometry, and an associated range of control parameters exhibiting normal
diffusion, with stronger topological instabilities. This way, we intend to
learn more about the emergence of possible fractal structures for diffusion
coefficients in billiards. As we will show, our model indeed generates a
considerably stronger irregular parameter dependence of the diffusion
coefficient than in the standard Lorentz gas. By applying the set of
approximation methods mentioned above we argue that long-range dynamical
correlations, or memory effects of orbits, are again at the origin of this
irregularity, as in case of simple one- and two-dimensional maps.

Our paper is composed of seven sections: In section \ref{model}, we introduce
the flower shape billiard.  Numerical results depicting the non-trivial
parameter dependence of the diffusion coefficient are shown in section
\ref{param}.  In section \ref{MZ},\ref{KD}, and \ref{GK}, we briefly review
the different approaches to understand the parameter dependence of diffusion
coefficients in deterministic dynamical systems, i.e., the Machta-Zwanzig
approximation, Klages-Dellago correction methods, as well as the approach
based on a suitable Green-Kubo formula for diffusion, and we apply them to the
flower shape billiard.  Summary and conclusions are contained in section
\ref{Summary}.

\section{The flower shape billiard}
\label{model}

The two-dimensional class of billiards we consider here consists of a point
particle of mass $m$ moving in a plane such that its Hamiltonian is
\begin{equation}
H = \frac{1}{2m}p_x^2+\frac{1}{2m}p_y^2 \: , \label{(1)}
\end{equation}
where $x$ and $y$ denote the Cartesian coordinates of the position in the
plane while $p_x$ and $p_y$ are the corresponding momenta.  The point particle
undergoes elastic collisions with obstacles that are fixed in the plane.  All
the obstacles have the same shape, and their centers are situated on a
triangular lattice according to
\begin{equation}
\qq_c = m_c \lla + n_c \llb\:, \label{(2)}
\end{equation}
as defined in terms of the fundamental translation vectors of the triangular 
lattice
\begin{equation}
\lla=(0,1)  \label{(3)}
\end{equation}
and
\begin{equation}
\llb=\left( \frac{\sqrt{3}}{2},\frac{1}{2} \right) \: , \label{(4)}
\end{equation}
where $m_c$ and $n_c$ are integers.

If all the pairs of integers are selected, we fill the whole triangular
lattice with hard wall obstacles, and the billiard is invariant under the
group of spatial translations generated by the vectors Eq.~(2). Accordingly,
the whole lattice can be mapped onto a so-called Wigner-Seitz cell with
periodic boundary conditions. The elementary Wigner-Seitz cell of the
triangular lattice is a hexagon of area
\begin{equation}
A_{WS}= |\lla \times \llb| = \frac{\sqrt{3}}{2} \: . \label{(5)}
\end{equation}

\begin{figure}
\psfig{figure=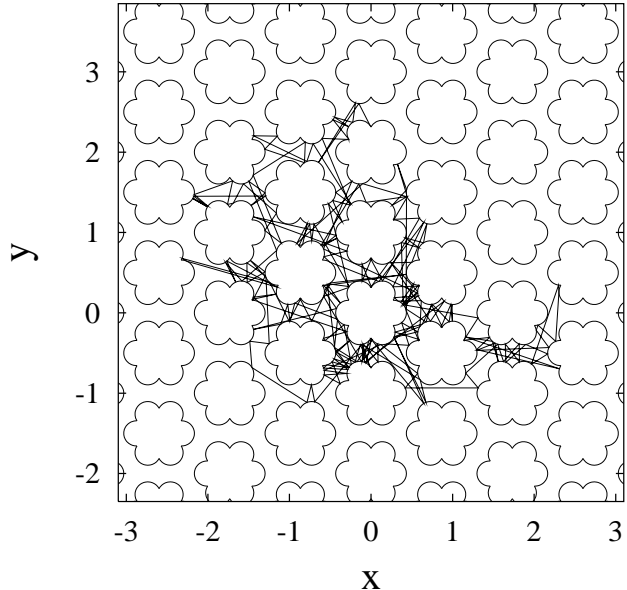,width=12cm} 
  
\vspace*{0.1cm}
\caption{The modified Lorentz gas as composed of a point particle moving
freely in the spaces between the flower-shaped obstacles, which scatters
elastically with the obstacles. In our case, mass $m=1$ and velocity
$v=1$. All figures are without units.}
\label{Fig1}
\end{figure}

In this paper, we propose an open billiard consisting of flower shape
obstacles instead of disks, which belongs to the general class of periodic
Lorentz gases whose normal diffusion has been proven by Bunimovich and Sinai
\cite{SinaiBunim}.  The mixing property and the extension of such billiards 
to higher-dimensional gases have been studied by Chernov
\cite{Chernov}.  As shown in Fig.~\ref{Fig1}, the space between the obstacles
forms the two-dimensional domain of the billiard where the point particle
moves freely and collides with the obstacles obeying the law of elastic
reflection.

\begin{figure}
\centerline{\psfig{figure=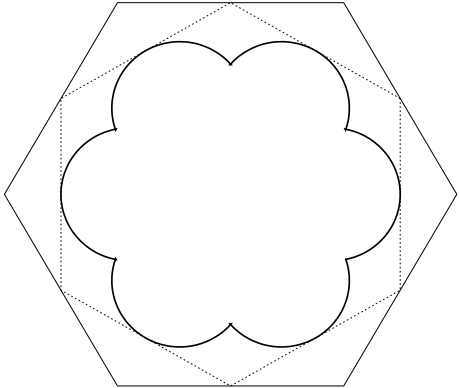,width=8cm}}
\caption{Definition of a flower-shaped obstacle. The bigger 
hexagon (bold lines) is the elementary Wigner-Seitz cell.  The arc always
touches the smaller hexagon (dotted lines), which prohibits any infinite
horizon.}
\label{Fig2}
\end{figure}

A single scatterer of our billiard is defined as follows: First, we consider
the inner hexagon whose vertices are on the middle points of the sides of the
hexagon of the elementary Wigner-Seitz cell, as depicted by the dotted lines
in Fig.~\ref{Fig2}.  Next, we join six arcs which have the same radii and
touch the inner hexagon.  Then we obtain the flower-shaped obstacle shown in
Fig.~\ref{Fig2}.  Note that the radius $r$ of one arc which consists in a
petal of the flower-shaped obstacle can be changed from $1/(4\sqrt{3})$ to
infinity.  According to this construction, the position space forms a
two-dimensional torus.  The motion of the point particle in the infinite
lattice is unbounded so that transport by diffusion is {\em a priori}
possible.  Indeed, we will show that the diffusion of point particles in the
billiard of the flower-shaped obstacles is normal. When the dynamics is
reduced to the Wigner-Seitz cell, the position of the particle inside this
cell must be supplemented by a lattice vector of the type of Eq.~(2) in order
to determine the actual position of the particle in the infinite lattice. This
lattice vector changes in discrete steps at each crossing of the border of the
elementary Wigner-Seitz cell.

A billiard whose obstacles are disks, or, in higher dimensions, spheres, is
called a periodic Lorentz gas, and this model serves as a typical example for
studying deterministic diffusion \cite{SinaiBunim,MZ,Chernov,GaBa95}.  The
diffusion coefficient of this standard periodic Lorentz gas has been studied 
in various ways both analytically and numerically, 
where recent work focused particularly
onto its density dependence (see Refs.\ \cite{KlagesDellago,KlaKor} and
further references therein). However, the density of this model cannot be
varied much because of the condition of finite horizon, which prohibits
collision-free ballistic motion and thus guarantees the existence of normal
diffusion \cite{SinaiBunim}. Consequently, the diffusion coefficient exists in
a very limited range of parameters only, and whether the diffusion coefficient
of the standard periodic Lorentz gas is a fractal function of the density of 
scatterers appears to be an open question.

Let us introduce the Liouville equilibrium invariant measure given by 
\begin{equation}
d\mu _e = I(x,y) \delta(H-E) dx dy dp_x dp_y, \label{(6)}
\end{equation}
where $I(x,y)$ is the indicator function of the billiard domain, and $E$ is
the energy of the point particle.  Averages over this invariant measure are
denoted by $<\cdot>$.  This measure is normalizable for the reduced dynamics
in an elementary Wigner-Seitz cell where the area of the billiard domain takes
a finite value. In this finite case, the Liouville invariant measure is a
probability measure which defines the microcanonical ensemble of equilibrium
statistical mechanics. The flower shape billiard belongs to the class of
dispersing billiards whose hyperbolicity has been proven by Sinai
\cite{Sinai}.  Consequently, it is known that the motion of the point particle
in the elementary Wigner-Seitz cell of our billiard is hyperbolic in the sense
that all orbits are unstable of saddle type with nonvanishing Lyapunov
exponents, and time averages are equal to averages over the Liouville
equilibrium invariant measure.

\section{ Curvature dependence of the diffusion coefficient}
\label{param}

Since the system of flower-shaped obstacles is fully chaotic, and by working
in the regime of finite horizon, we may expect that diffusion is normal in the
sense that the position is asymptotically a Gaussian random variable with a
variance growing linearly in time. Consequently, the diffusion coefficient
exists and is finite \cite{SinaiBunim,Chernov}.  Indeed, we checked
numerically that the variance is proportional to time after sufficiently long
time evolution.

\begin{figure}
\centerline{\psfig{figure=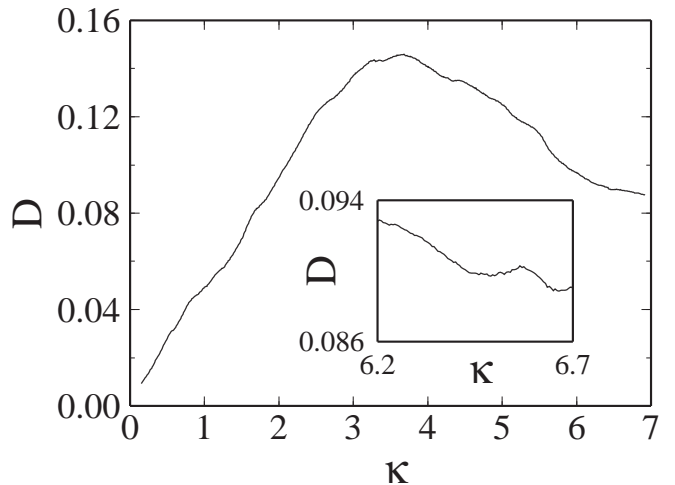,width=8.2cm}}
  
\vspace*{0.1cm}
\caption{Diffusion coefficient $D$ (solid line) versus the curvature 
$\kappa$ of the petal of the flower-shaped obstacles.  The diffusion
coefficient inceases approximately linearly for small enough $\kappa$ until
reaching a global maximum.  Inset: Zoom on the curve of the diffusion
coefficient for larger $\kappa$ showing the irregularity of this curve on fine
scales.}
\label{mlgdif}
\end{figure}

The diffusion coefficient $D$ is given by the Einstein formula,
\begin{equation}
D = \lim_{t\rightarrow\infty}\frac{1}{4t}
<\{\qq(t)-\qq(0)\}^2>\:,
\label{(9)}
\end{equation}

\noindent
and according to this formula the diffusion coefficient was calculated from
computer simulations in the flower shape billiard 
where the curvature $\kappa$ of the petals is varied 
from 0 to its maximum $4\sqrt{3}$. 
The results are depicted in Fig.~\ref{mlgdif}.  
In this figure, we observe a non-trivial structure
depending on the curvature $\kappa$ of the arc defining the petal of 
the flower-shaped obstacles.

The gross features of the curvature dependence for the diffusion
coefficient can qualitatively be explained as follows: When the curvature of
the petal of the flower-shaped obstacle is zero, the inner hexagon shown by
the dotted lines in Fig.~\ref{Fig2} connects to the six hexagons surrounding
it.  In this case, the point particle remains forever localized in compact
domains bounded by the three neighbouring hexagons. For this specific value of
the control parameter, the motion of the point particle is completely
predictable because the compact domain is an equilateral triangle, and the
system is integrable.

When the curvature becomes positive, the point particle can run away from the
compact domain, and diffusion occurs. As already explained, at all positive
curvatures of the petal, even if they are very small, the motion of the point
particle is fully chaotic and the horizon is finite, 
hence diffusion is expected to be normal. The
diffusion coefficient starts to increase from zero according to the linear
increase of the curvature of the petal, and related to the fact that the space
between petals also increases.

When the radius of the petal is equal to $R_L=\sqrt{3}/4\simeq0.433$, which is
the distance between the center of the hexagon and the tangent point to the
hexagon, the obstacle becomes a disk of radius $R_L$, that is, for this
parameter value our billiard is precisely the same as the conventional
periodic Lorentz gas. This point corresponds to the curvature $\kappa\simeq
2.309$ in Fig.~\ref{Fig2}.

When the radius $r$ of the curvature of the petal decreases below $R_L$, the
point particle is much more likely to be trapped in the space between two
obstacles. 
This appears to be due to the formation of wedges between any two
petals of a flower shape obstacle.

The inset of Fig.~{\ref{mlgdif}} depicts a zoom on the curve showing the fine
structure on smaller scales with respect to curvature.  We remark that the
apparently continuous fluctuations therein are within the numerical errors,
that is, we confirmed the convergence of our results within a precision of
order $10^{-4}$ by taking an average over $10^{10}$ initial conditions.
Unfortunately, with our computational power it is impossible to check whether
this oscillatory behavior persists on even finer scales.

\section{ Machta-Zwanzig approximation for diffusion coefficients }
\label{MZ}

In Ref.\ \cite{MZ}, Machta and Zwanzig have obtained a simple analytical
approximation for the diffusion coefficient of the periodic Lorentz gas which
yields asymptotically correct results in the limit of small gaps between
disks. In this case, the particle is for a long time somewhat trapped in the
triangular regions between three adjacent scatterers. Hence, the particle is
supposed to loose the memory of its past itinerary due to the multiple
scattering in the trap region, and the transition probabilities to the
neighbouring triangular cells are assumed to be equivalent. As was shown in
Ref.\ \cite{MZ}, the average rate $\tau ^{-1}$ of such transitions can be
calculated from the fraction of phase space available for leaving the trap
divided by the total phase space volume of the trap leading to
\begin{equation}
\tau=\pi A/(3W) \label{tau} \:,
\end{equation}
where $A$ is the area of the trap and $W$ the width of the gap between the
disks.

The flower shape billiard has similar types of traps as the periodic Lorentz
gas.  Accordingly, the Machta-Zwanzig approximation can be applied to the
flower shape billiard as well, and Eq.~(\ref{tau}) holds again for the average
trapping time. Hence, we only need to calculate the area of the trap and the
gap between the petals from simple geometrical considerations yielding
\begin{equation}
A=\frac{3\sqrt{3}}{4}-3h\left[ \sqrt{3}h+\sqrt{r^2-h^2} \right]
\label{(10)}
\end{equation}
and 
\begin{equation}
W=\frac{1}{2}-\left[ \sqrt{3}h+\sqrt{r^2-h^2} \right] \: ,
\label{(11)}
\end{equation}
 
\noindent
where 
\begin{equation}
h=\frac{1}{2}\left( \frac{\sqrt{3}}{4}-r \right).
\label{(12)}
\end{equation}
In the above, $r$ denotes the radius of the curvature of the petal.

\begin{figure}
\centerline{\psfig{figure=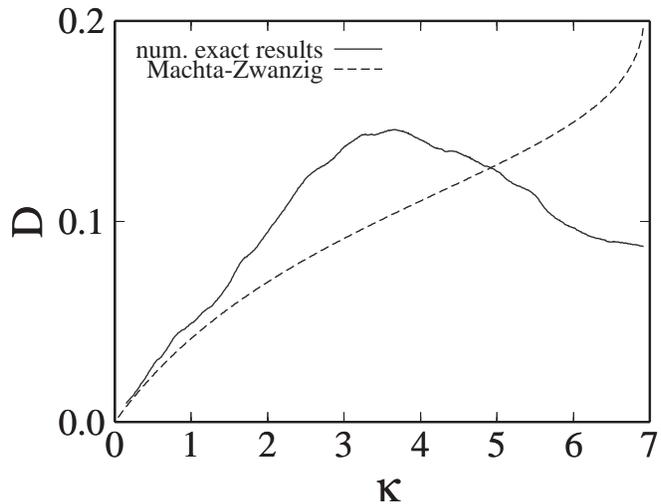,width=8.2cm}}
 
\vspace*{0.1cm}
\caption{Diffusion coefficient $D$ (solid line) versus the curvature 
$\kappa$ of the petal of the flower-shaped obstacle.  The solid curve
correspond to the numerically exact results while the dotted curve yields the
Machta-Zwanzig random walk approximation Eq.~(\ref{(13)}).}
\label{mz}
\end{figure}

\noindent
The distance $l$ between the centers of the flower shape obstacles is
$1/\sqrt{3}$. Assuming that the gap size $W$ is very narrow leads to the
Machta-Zwanzig random walk approximation for the diffusion coefficient
\begin{equation}
D_{MZ}=\frac{l^2}{4\tau}
\label{(13)}
\end{equation}
with $\tau$ being given by Eq.~(\ref{tau}) and supplemented by
Eqs.~(\ref{(10)})-(\ref{(12)}) for the flower shape case. As is shown in
Fig.~\ref{mz}, the Machta-Zwanzig approximation works very well in the
vicinity of zero curvature of the petal only.

\section{ Klages-Dellago corrections of the
Machta-Zwanzig approximation }
\label{KD}

In Ref.\ \cite{KlagesDellago}, Klages and Dellago have generalized 
the Machta-Zwanzig approximation for the standard periodic Lorentz gas 
by taking memory effects of orbits into account. 
Their generalization is based on the observation that, except in 
the asymptotic limit of narrow gap sizes, the diffusive dynamics is not 
a simple Markov process, in the sense that there
exist non-vanishing dynamical correlations. By mapping the orbit of a particle
onto a suitable symbolic dynamics they numerically calculated the
probabilities to obtain certain symbol sequences of finite length. Increasing
the length of these symbol sequences yielded systematic corrections of the
Machta-Zwanzig approximation. In Ref.\ \cite{KlagesDellago}, two schemes
directly emerging from this approach were discussed, one suggesting simple
heuristic corrections to the simple random walk model of 
diffusion Eq.~(\ref{(9)}),
and another one employing lattice gas computer simulations defined by these
probabilities. In this section, we apply these two methods to the flower shape
billiard in order to systematically correct the Machta-Zwanzig approximation.
A third scheme starting from a Green-Kubo formula for diffusion will be
discussed in section \ref{GK}.

\begin{figure}
\centerline{\psfig{figure=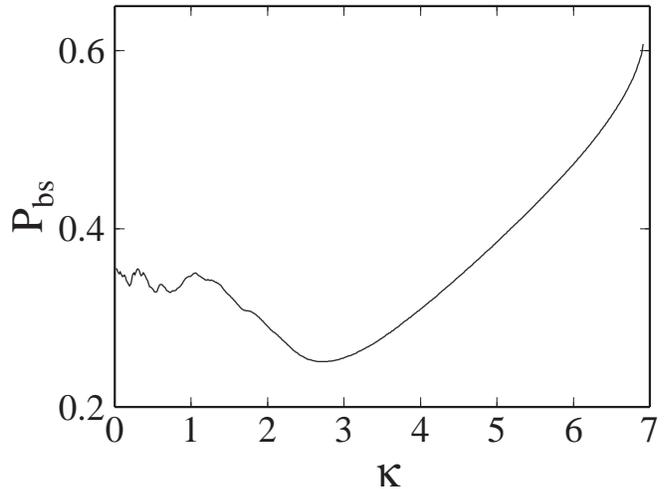,width=8.2cm}}
  
\vspace*{0.1cm}
\caption{Backscattering probability $p(z)$ (solid line) versus the curvature 
$\kappa$ of the petal of the flower-shaped obstacle.  In the case of a
Markovian process it is equal to $1/3$.}
\label{bs}
\end{figure}

The Machta-Zwanzig approximation assumes that a particle jumps from one trap
to a neighbouring trap situated on the hexagonal lattice of traps. However,
there exist non-vanishing probabilities that a particle can jump to next
nearest neighbours, or even farther, without collisions. Accordingly, we
should correct the Machta-Zwanzig approximation for the flower shape billiard
by using the probabilities $p_{cf1}$ and $p_{cf2}$ of those collisionless
flights which lead from one cell directly to its second nearest neighbours, or
to the third nearest neighbours, respectively. The distances $l_1$ and $l_2$
between the centers of a trap to respective second and third neighbours are
\begin{equation}
l_1=\sqrt{3}l\:, \qquad l_2=\sqrt{7}l \:.
\label{(14)}
\end{equation}
The diffusion coefficient $D_{cf}$ with corrections due to these collisionless
flights then reads
\begin{eqnarray}
D_{cf}&=&(1-p_{cf1}-p_{cf2})D_{MZ}+p_{cf1}\frac{l_1^2}{4\tau}
+p_{cf2}\frac{l_2^2}{4\tau} \nonumber \\ 
&=&(1+2p_{cf1}+6p_{cf2})D_{MZ} \:.
\label{(15)}
\end{eqnarray}  
Next we take memory effects of orbits due to backscattering into account.  For
this purpose, orbits are coded by labeling the entrance through which a
particle enters a trap with $z$, the exit to the left of this entrance with
$l$, and the one to the right with $r$. Thus, an orbit can be mapped onto a
sequence of symbols $z,l$, and $r$.  For example, $p(z)$ is the backscattering
probability $p_{bs}$, which is the probability of the moving particle to leave
the trap through the same gate where it entered. The Machta-Zwanzig
approximation assumes that $p(z)=p(l)=p(r)=1/3$. However, in general $p(z)$ is
not close to $1/3$ as shown in Fig.~\ref{bs}, because the actual orbits do not
loose their memory during their itineraries.  

A more profound explanation for the
complicated functional form of $p(z)$ may be provided in terms of the theory
of chaotic scattering: Chaotic scattering systems with multiple exit modes
typically have fractal phase space boundaries separating the sets of initial
conditions (basins) going to the various exits. However, open systems such as
a three-disk scatterer of the periodic Lorentz gas possess the even stronger
property of being {\em Wada}, that is, any initial condition which is on the
boundary of one exit basin is also simultaneously on the boundary of all the
other exit basins \cite{PCOG96}. Changing the curvature $\kappa$ sensitively
affects the highly irregular structure of these basin
boundaries. Consequently, Fig.~\ref{bs} may be understood as reflecting the
topological instability of Wada basins under parameter variation, and as we
will now show this is reflected in the parameter dependence of the diffusion
coefficient.

Modifying the Machta-Zwanzig random walk by including the backscattering
probability $p(z)$ we obtain the diffusion coefficient
\begin{equation}
D_{BS}=\frac{(1-p(z))l_2^2}{4(2\tau)}=(1-p(z))\frac{3}{2}D_{MZ}\:.
\label{(16)}
\end{equation}  

\noindent
Combining the effects of collisionless flights and backscattering yields as a
first order approximation
\begin{equation}
D_1=\frac{3}{2}(1-p(z))(1+2p_{cf1}+6p_{cf2})D_{MZ}\:.
\label{(17)}
\end{equation}

\begin{figure}
\centerline{\psfig{figure=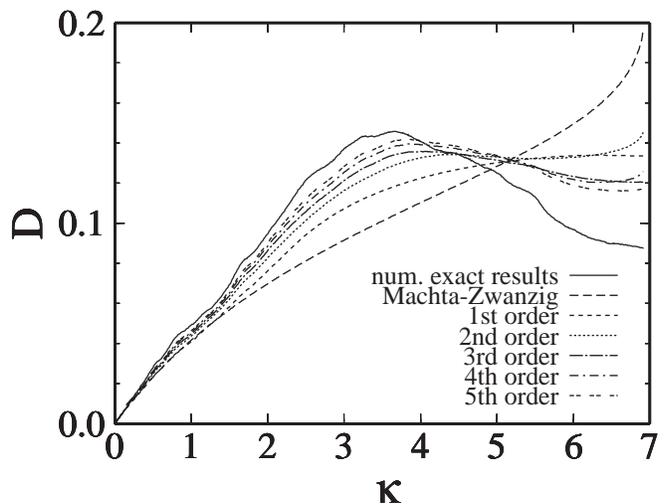,width=8.2cm}}
  
\vspace*{0.2cm}
\caption{Diffusion coefficients of higher-order approximations due to
including higher-ordering backscattering probabilities. The solid curve
corresponds to numerically exact results while the other curves represent
approximate solutions.}
\label{high}
\end{figure}

\noindent
Higher-order approximations of the diffusion coefficient, as related to longer
symbol sequences and respective probabilities such as $p(lrz\ldots)$, can be
derived in the same way \cite{KlagesDellago}. For the flower shape billiard,
respective results are shown in Fig.~\ref{high}.

The above correction methods assume that all orbits follow higher-order
Markov processes, where correlations are present in form of initial transient
times before the variance becomes linear in time. This dynamics appears to be
more suitably represented in form of lattice gas simulations on a honeycomb
lattice, where the sites of the lattice represent the traps. Indeed, for the
periodic Lorentz gas such lattice gas simulations were performed in Ref.\
\cite{KlagesDellago} confirming fast convergence to the numerically exact
results. Compared to that scheme, the convergence of the intuitive correction
method described above is firstly slower, and secondly it is not everywhere
converging to the numerically exact results, which is due to the fact that
this approach was purely of a heuristic nature.

We also performed lattice gas simulation in case of the flower shape billiard
according to the following prescription: Particles 
hop from site to site with frequency $\tau ^{-1}$, which is 
identical to the hopping frequency used in the Machta-Zwanzig
approximation. The hopping probabilities are given by the backscattering
probability $p(z)$ and by those corresponding to respective longer symbol
sequences.  The diffusion coefficient is then obtained from the Einstein
formula Eq.\ (\ref{(9)}) in the limit when the variance is getting
proportional to time. The correlations in the actual orbits are thus
systematically and exactly filtered out according to the length of the symbol
sequences. 

\begin{figure}
\centerline{\psfig{figure=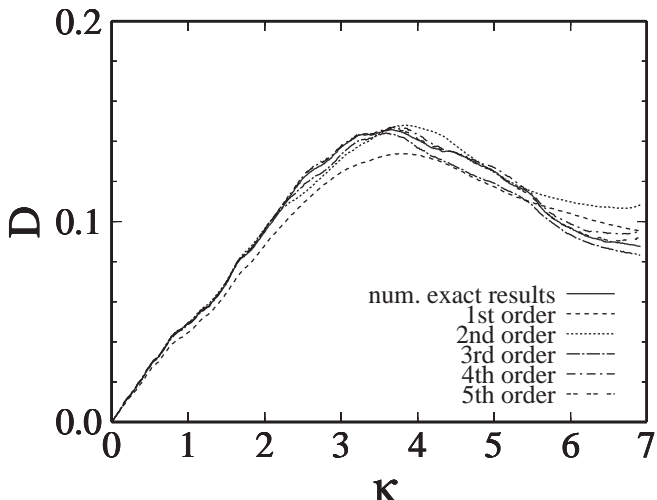,width=8.2cm}}
  
\vspace*{0.1cm}
\caption{Diffusion coefficient as obtained from lattice gas simulations based
on higher-order backscattering probabilities. The solid curve corresponds to
numerically exact results while the other curves yield higher-order
approximations.}
\label{lg}
\end{figure}

In Fig.~\ref{lg}, the results of such higher-order approximations according to
lattice gas simulations are shown. One can see that the convergence to the
numerically exact results is not only much better than in Fig.~\ref{high}, but
even exact. Strong memory effects are clearly visible especially after the
diffusion coefficient curve takes its maximum. In the previous heuristic
modifications to the simple random walk model, 
the dynamics was only modeled for a limited number of time steps 
as a Markov process. Fig.~\ref{high} suggests
that correlations as contained in the symbol sequences are more suitably
represented by higher-order iterations in form of lattice gas simulations.
However, a disadvantage is that the lattice gas scheme requires a second round
of computations which is put on top of the previous simulations, by again
looking at the time evolution of an initial ensemble of points.

\section{ The Green-Kubo formula approach }
\label{GK}

The main drawbacks of the two methods described above were, firstly, that the
heuristic corrections of the Einstein formula were not converging exactly to
the numerically exact results, and, secondly, that the lattice gas simulations
were merely a numerical scheme without being represented in form of analytical
approximations. These deficiencies were essentially resolved in Ref.\
\cite{KlaKor} by the derivation of a Green-Kubo formula which employs the
symbolic dynamics on the hexagonal lattice of traps introduced in
section~\ref{KD}. The result reads
\begin{equation}
D= \frac {1}{4\tau}C_0+\frac{1}{2\tau}\sum_{n=1}^\infty C_n \quad ,
\label{eq:gk2}
\end{equation}
with
\begin{equation}
C_n:=\left\langle {\bf j}({\bf x}_0)\cdot{\bf j}({\bf x}_n)\right\rangle \label{eq:vacf}
\end{equation}
being the velocity autocorrelation function related to jumps ${\bf j}({\bf
x}_n)$ on the hexagonal lattice at time step $n$. These jumps are suitably
defined in terms of the lattice vectors Eqs.~(\ref{(3)}),(\ref{(4)}). That is,
any symbol sequence of an orbit on the hexagonal lattice of traps defines a
respective chain of lattice vectors. The averages indicated by the brackets in
Eq.\ (\ref{eq:vacf}) are calculated by weighting the respective scalar
products of lattice vectors with the corresponding conditional probablities
$p(\alpha\beta\gamma\ldots)\;,\;\alpha,\beta,\gamma\in\{l,r,z\}$.  
In Eq.\ (\ref{eq:vacf}), $\tau$ is the mean time of free flight 
between symbol changes, and it is given by Eq. (\ref{tau}).
Eq.\ (\ref{eq:gk2}) is thus the honeycomb lattice analogue to the Green-Kubo
formula derived by Gaspard for the Poincar\'e-Birkhoff map of the periodic
Lorentz gas \cite{GaspardCUP,Gasp96}. 

It is easy to see that the first term in Eq.\ (\ref{eq:gk2}) yields the
Machta-Zwanzig approximation Eq.~(\ref{(13)}). Higher-order corrections can
then be calculated by defining the hierarchy of approximations
\begin{equation}
D_n=
\frac{l^2}{4\tau}+\frac{1}{2\tau}\sum_{\alpha\beta\gamma\ldots}p(\alpha\beta\gamma\ldots)
\bml{\ell}\cdot\bml{\ell}(\alpha\beta\gamma\ldots) \label{eq:gka3}
\end{equation}
with $n>0$ being the number of symbols and $D_0(w)$ given by Eq.\
(\ref{(13)}), where, again, $\bml{\ell}(\alpha\beta\gamma\ldots)$ are suitable
lattice vectors.

\begin{figure}
\centerline{\psfig{figure=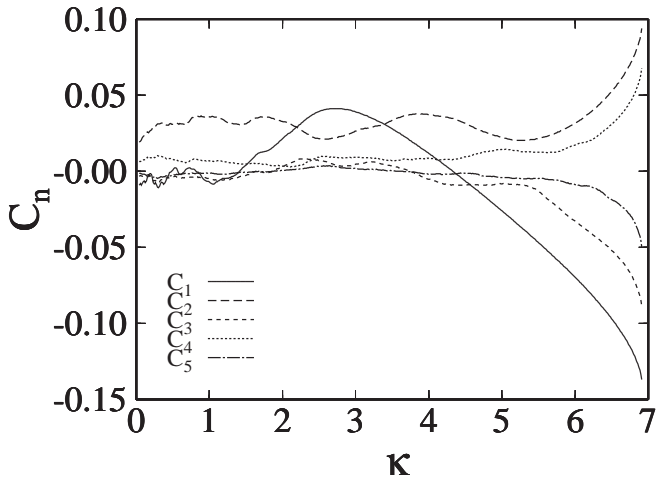,width=8.2cm}}
  
\vspace*{0.1cm}
\caption{Parameter dependence of the time-dependent correlation function 
$C_n$, see Eq.\ (\ref{eq:vacf}), 
as defined with respect to the symbolic dynamics on the hexagonal lattice of
traps. At any parameter $C_n$ decays
exponentially related to the fact that the Green-Kubo formula Eq.\
(\ref{eq:gk2}) is a convergent series. The speed of the convergence depends on
the curvature. Obviously, in the large curvature region the correlation
function decays more slowly than for small curvature.}
\label{cn}
\end{figure}

\begin{figure}
\centerline{\psfig{figure=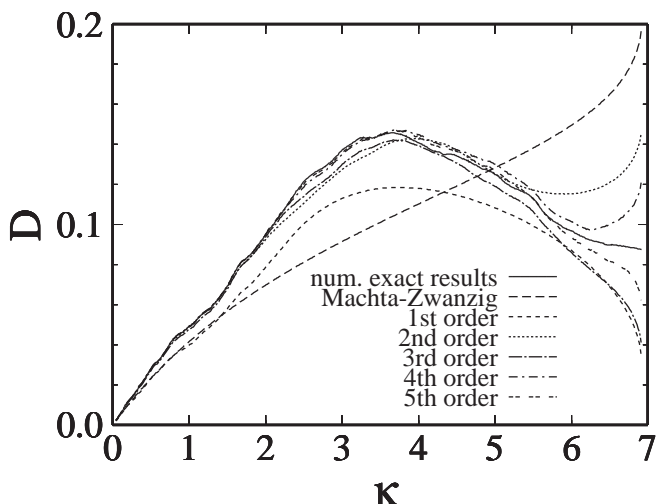,width=8.2cm}}
  
\vspace*{0.1cm}
\caption{Diffusion coefficients as obtained from the Green-Kubo formula
Eq.\ (\ref{eq:gka3}).  The solid curve correspond to the asymptotic,
numerically exact results while the other curves yield the respective
hierarchy of approximations.}
\label{gk}
\end{figure}

The impact of dynamical correlations on the diffusion coefficient can now be
understood by analyzing the single contributions in terms of the correlation
function $C_n$ as contained in the Green-Kubo formula Eq.~(\ref{eq:gk2}). In
fully chaotic systems such as the periodic Lorentz gas and the flower shape
billiard, the velocity correlation function decays exponentially, which is in
agreement to the results depicted in Fig.~\ref{cn}. By comparing this figure
to Fig.~\ref{gk} one can learn how the irregularities of the correlation
function determine the parameter dependent diffusion coefficient: Let us start
with the first-order approximation of Eq.~(\ref{eq:gka3}) which reads
$D_1=D_0+D_0(1-3p(z))$. The functional form of $p(z)$ in Fig.~\ref{bs} thus
qualitatively explains the position of the global maximum of the diffusion
coefficient, because at this value of the curvature the probability of
backscattering is minimal. Adding up the three-jump contributions coming from
$C_2$ furthermore yields the most important quantitative contributions in this
region of the curvature. In the region of large curvature the diffusion
coefficient decays monotonically according to the effect of two-hop
correlations covered by $C_1$. However, note the large fluctuations of the
correlation function $C_n$ as well as of the diffusion coefficient
approximations $D_n$ in this regime both indicating the dominant effect of
long-range higher-order correlations. Studying the detailed convergence of the
approximations depicted in Fig.~\ref{gk} shows that correlations due to orbits
with longer symbol sequences yield irregularities in the parameter dependence
of the diffusion coefficients on finer and finer scales.

\section{ Summary and conclusion }
\label{Summary}

In this paper, we have introduced a novel variant of the periodic Lorentz gas
by assigning a flower-shaped geometry to the scatterers. Although both systems
are rather similar in the sense that they are both fully chaotic and exhibit
normal diffusion in a certain parameter range, we have found that the
diffusion coefficient in the flower shape geometry is considerably more
irregular under parameter variation than the one obtained from circular disks
as scatterers. We have analyzed these irregularities by three different
methods which all start from correcting the Machta-Zwanzig random walk
approximation for the diffusion coefficient. All these improved approximation
schemes use a symbolic dynamics which maps the orbits of moving particles to
symbol sequences according to traps situated on a hexagonal lattice. We have
discussed the convergence of these different approximation schemes, and we
have shown how they enable a detailed understanding of the precise shape of
the parameter dependent diffusion coefficient in the flower shape billiard in
terms of long-range dynamical correlations.

The Green-Kubo formula introduced in Ref.\ \cite{KlaKor} appears to be most
suitable for understanding the irregular behavior of the parameter dependent
diffusion coefficient, because it conveniently transforms the diffusive
dynamics into a sum over the velocity correlation function, whose specific
parameter dependence can in turn be analyzed step by step. In particular, this
approach yields an exact convergence to the parameter dependent diffusion
coefficient as obtained from simulations.

Interestingly, when the correlation function decays in time, the frequency of
oscillations as a function of the control parameter increases. The relation
between this decay in time and the increase of the frequency of these
oscillations determines the strength of the irregularities on fine scales of
the resulting parameter dependent diffusion coefficient. The question of the
existence of fractal diffusion coefficients in billiards such as periodic
Lorentz gases with circular or flower-shaped scatterers might thus be answered
by using Green-Kubo formulas if the respective correlation functions could be
evaluated more precisely for large enough times. Indeed, in Ref.\
\cite{HarayamaGaspard} the highly irregular diffusion coefficient of an open
billiard in an external field has already been investigated along these lines
by relating the Poincar\'{e}-Birkhoff version of the Green-Kubo formula to
fractal Weierstrass functions. The joint efforts compiled in Refs.\
\cite{KlagesDellago,HarayamaGaspard,KlaKor} may therefore be considered as
first steps towards answering the conjecture of Refs.\
\cite{KlagesDorfman,Klages}, which suggested a possible universality of
fractal diffusion coefficients in low-dimensional fully chaotic dynamical
systems exhibiting some spatial periodicity, for the case of chaotic 
Hamiltonian dynamical systems such as particle billiards.


\acknowledgements
T.~H. thanks Dr. B. Komiyama for support 
and encouragement in this research. 
R.~K. thanks Mr. J.~Aguirre for pointing out Refs.~\cite{PCOG96} to him. 
P.~G. thanks the National Fund for Scientific Research (F.N.R.S. Belgium) 
for financial support.


\end{document}